\documentclass[prd,floats,superscriptaddress,showpacs,nofootinbib,twocolumn,preprintnumbers]{revtex4}

\usepackage{amssymb}
\usepackage{amsmath}

\usepackage{sans}

\usepackage{graphicx}
\usepackage{graphics}
\usepackage{dcolumn}
\usepackage{color}
\usepackage{rotate}
\usepackage{fancyhdr} 
\usepackage{hyperref}
\definecolor{azure(colorwheel)}{rgb}{0.0, 0.5, 1.0}
\hypersetup{colorlinks=true,linkcolor=black,citecolor=azure(colorwheel),pdfborder={0 0 0}}
\usepackage{indentfirst}
\usepackage{fontawesome}
\usepackage{hyperref}
\usepackage{umoline}

\def\be{\begin{eqnarray}}
\def\ee{\end{eqnarray}}
\def\ba{\begin{eqnarray}}
\def\ea{\end{eqnarray}}
\def\no{\nonumber}

\definecolor{darkred}{rgb}{.743,0,0}

\newcommand{\refeq}[1]{Eq.~(\ref{eq:#1})}          
          
\newcommand{\reffig}[1]{Fig.~\ref{fig:#1}}

\usepackage{xspace}
\newcommand{\BarA}{Bar18\xspace}
\newcommand{\BarB}{Bar19\xspace}

\begin{document}
\title{Galactic rotation curves versus ultralight dark matter: A systematic comparison with SPARC data}

\author{Nitsan Bar} \email{nitsan.bar@weizmann.ac.il}
\affiliation{Department of Particle Physics and Astrophysics,
	Weizmann Institute of Science, Rehovot 7610001, Israel}
  
\author{Kfir Blum} \email{kfir.blum@weizmann.ac.il}
\affiliation{Department of Particle Physics and Astrophysics,
 	Weizmann Institute of Science, Rehovot 7610001, Israel}
  
\author{Chen Sun}\email{chensun@mail.tau.ac.il}
\affiliation{School of Physics and Astronomy, Tel-Aviv University, Tel-Aviv 69978, Israel}

\date{\today}

\begin{abstract}
We look for and place observational constraints on the imprint of ultralight dark matter (ULDM) soliton cores in rotation-dominated galaxies. Extending previous analyses, we find a conservative constraint which disfavors the soliton-host halo relation found in some numerical simulations over a broad range in the ULDM particle mass $m$. Combining the observational constraints with theoretical arguments for the efficiency of soliton formation via gravitational dynamical relaxation, and assuming that the soliton-halo relation is correct, our results disfavor ULDM from comprising 100\% of the total cosmological dark matter in the range $10^{-24}~{\rm eV}\lesssim m\lesssim10^{-20}~{\rm eV}$. The constraints probe the ULDM fraction down to $f\lesssim0.3$ of the total dark matter. 
\href{https://github.com/ChenSun-Phys/ULDM_x_SPARC}{\faGithub}
\end{abstract}


\maketitle

 \tableofcontents


\section{Introduction and main result}
\label{sec:introduction}

Ultralight bosonic fields offer a plausible candidate for dark matter (DM). The wave nature of such ultralight dark matter (ULDM) may manifest itself in a variety of astrophysical settings
~
\cite{Preskill:1982cy,Abbott:1982af,Dine:1982ah,
  Blas:2016ddr,
  Kendall:2019fep,
  Poddar:2021sbc,
  KumarPoddar:2019jxe,
  Dror:2020zru,
  Agrawal:2017cmd}. 
A wide range of the particle mass, $ 10^{-25}\lesssim m \lesssim 10^{-19} $~eV, can be probed via observations of galaxies~
\cite{Schutz:2020jox,Amorisco:2018dcn,Lora:2011yc,Lancaster2020,Bar-Or:2020tys,Dalal:2020mjw,Deng:2018jjz,Guo:2020tla,Eby:2020eas,Broadhurst:2018fei,Church:2018sro}
(for reviews, see~\cite{Niemeyer:2019aqm,Hui:2021tkt}). At the lower end in $m$, Ref.~\cite{Blum:2021oxj} argued that strong gravitational lensing by massive elliptical galaxies is sensitive to ULDM as light as $ m\sim 10^{-25} $~eV, making up a small fraction of the order of 10\% of the total cosmological DM. At the higher end, Ref.~\cite{Marsh:2018zyw} suggested that ULDM-induced dynamical heating in small satellite galaxies may probe $ m\sim 10^{-19} $~eV. Many studies highlighted $ m\sim 10^{-22} $~eV, for which ULDM was suggested as a solution to small-scale puzzles facing cold dark matter (CDM) \cite{Hu:2000ke,Schive:2014dra,Schive:2014hza,Hui:2016ltb}. However, with further scrutiny, this proposal became increasingly implausible. The possibility that ULDM at $m\sim10^{-22}$~eV comprises the majority of the DM is in tension with Lyman-$ \alpha $ forest~\cite{Irsic:2017yje,Kobayashi:2017jcf,Armengaud:2017nkf,Zhang:2017chj,Nori:2018pka,Rogers:2020ltq} and cosmic microwave background anisotropy analyses~\cite{Hlozek:2017zzf,Lague:2021frh}, as well as with stellar and gas kinematics in low-surface-brightness galaxies~\cite{Bar:2018acw,Bar:2019bqz} and dwarf galaxies~\cite{Safarzadeh:2019sre} (see also~\cite{Bernal:2017oih,Robles:2018fur} for related analysis).

In this paper we concentrate further on galactic signatures of ULDM. An important prediction, observed in numerical simulations~\cite{Schive:2014dra,Schive:2014hza}, is the formation of a ``soliton'' density core in the halo  center. The soliton is a ground state configuration of the equations of motion. Reference~\cite{Schive:2014hza} found that the soliton mass in their simulations is related to the host halo via the so-called soliton-host halo relation. References~\cite{Bar:2018acw,Bar:2019bqz} (hereafter \BarA and \BarB, respectively) showed that the empirical soliton-host halo relation is equivalent to the equilibration of specific kinetic energy (kinetic energy per unit mass of the field) in the soliton and in the halo: $ K/M|_{\rm soliton} \approx K/M|_{\rm halo}$. While exact equilibration cannot be the end state of a self-gravitating system, the observed scaling is likely a bottleneck state to which the system is driven by gravitational dynamical relaxation~\cite{Hui:2016ltb,Levkov:2018kau,Veltmaat:2018dfz,Chavanis:2019faf,Eggemeier:2019jsu,Chen:2020cef,Schwabe:2020eac}.

\BarA used rotation curve data from the SPARC database~\cite{Lelli:2016zqa} to look for the imprint of the solitons predicted by the soliton-host halo relation. The result was null; thus, assuming the soliton-halo relation observed in simulations of Refs.~\cite{Schive:2014dra,Schive:2014hza} is correct, ULDM in the range $10^{-22}\lesssim m\lesssim 10^{-21} $~eV is disfavored by the data. 

We believe that this (unfortunately, null) result is significant: ULDM provided a theoretically plausible model of DM, for which the soliton-halo relation of Refs.~\cite{Schive:2014dra,Schive:2014hza} formed a sharp prediction of an observable feature, without invoking any interactions between DM and the Standard Model particles apart from minimal gravity alone. The implications of a positive detection of this feature in a variety of different galaxies could have been far reaching. The implications of not detecting the feature are also substantial, because they have the potential to exclude a whole swath of the mass range of DM.

\begin{figure}[th]
	\centering
	\includegraphics[width=.49\textwidth]{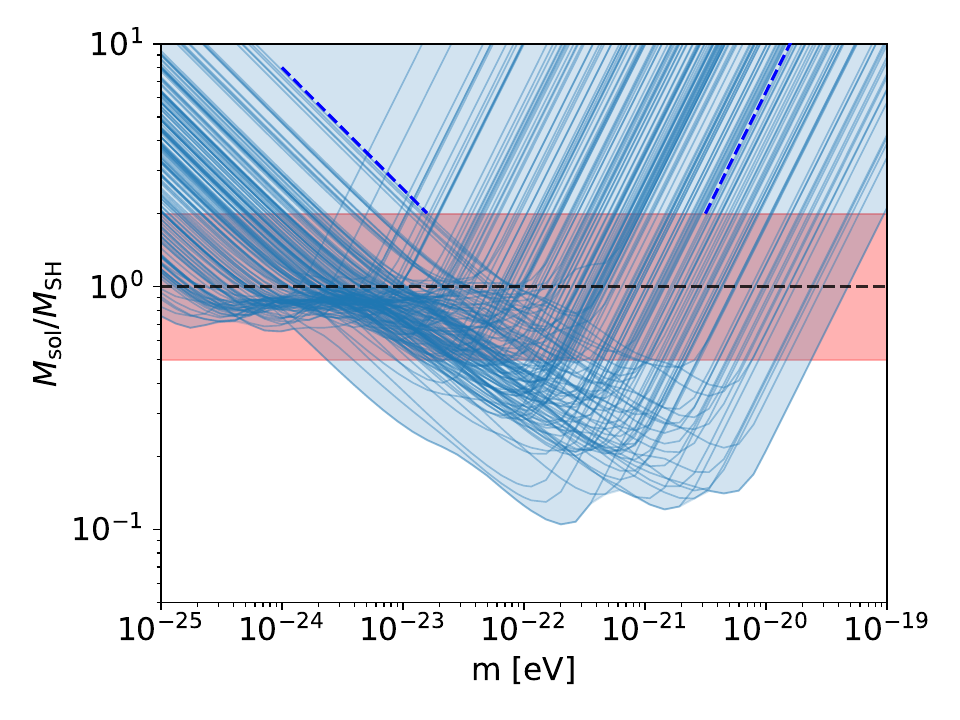}
	\caption{The combined 95\%~C.L. constraints (solid blue) of SPARC galaxies on the mass of the soliton $M$, normalized by that predicted by soliton-host halo relation, $M_{\rm SH}$. Each blue line corresponds to a galaxy. The blue dashed lines highlight analytical approximations, valid at small and large $m$. The red band comes from allowing $M_{\rm SH}$ to vary by up to a factor of 2 up or down. See Sec.~\ref{ssec:data} for more details of the computation.}
	\label{fig:final-result-intro}
\end{figure}

With this motivation in mind, in the current work we expand on \BarA in a number of aspects. 
First, whereas \BarA reported only a crude estimate of the observationally disfavored range in $m$, we perform a systematic scan of the data, resulting in broader and more comprehensive limits. 
A summary of our results is shown in Fig.~\ref{fig:final-result-intro}. The blue region combines the constraints from all of the SPARC galaxies; each thin line corresponds to a single individual galaxy. On the $y$~axis, we use the soliton mass $M$, normalized to the mass specified by the soliton-halo relation, $ M_{\rm SH} $. We allow an uncertainty of a factor of 2, up or down (cf. Ref.~\cite{Schive:2014hza} and \BarA), in $M_{\rm SH}$, represented by the red band. On the $x$~axis, we show the ULDM mass $m$. Where the blue region dips below the red band, which happens for $3\times 10^{-24} <m<2\times 10^{-20} $~eV, the soliton-halo relation is in conflict with the data. 
%
%

It is important to note that in many galaxies, the soliton-halo relation is {\it not generally expected to hold} for $m\gtrsim10^{-21}$~eV. The relation was tested only by numerical simulations (of galaxies similar to DM-dominated SPARC systems) for $m\sim10^{-22}$~eV; when $m$ is increased, the efficiency of dynamical relaxation diminishes, and eventually one expects the relation to break down because the soliton does not have enough time to form during the age of the galaxy. From this perspective, the $m\gtrsim10^{-21}$~eV part of the excluded range in Fig.~\ref{fig:final-result-intro} may not be very informative. Nevertheless, as we explain later on, the full excluded range (including the higher $m$ range) is still of phenomenological interest. First, it is an  observational constraint and can be considered as a null search for the soliton imprint, putting aside theoretical bias. The sensitivity that we demonstrate in the data strongly motivates additional simulation analyses to test the extension of the theoretical soliton-halo relation up to higher $m$.
Second, we shall see that in a class of models (inspired by the ``string axiverse" scenario~\cite{Arvanitaki:2009fg}) in which more than one species of ULDM coexist, dynamical relaxation could become unexpectedly efficient and populate the soliton state even for high-$m$ fields: In other words, there are interesting and well-motivated theoretical models that could be expected to produce a soliton even for $m>10^{-21}$~eV.

Another aspect which we extend over \BarA is to use the theoretical analysis of dynamical relaxation in order to derive constraints also in the case where ULDM comprises just a fraction $f<1$ of the total cosmological DM. 


An outline of the rest of this paper is as follows. In Sec.~\ref{sec:backgrounds}, after briefly reviewing the soliton--host halo relation, we derive observational constraints on this relation using SPARC data. Most of our results are model independent and conservative, in that we consider only a rotation curve to constrain a soliton feature if the soliton feature, by itself and without considering any additional mass component, overshoots the velocity data. In Sec.~\ref{sec:interpr-constr-solit} we use theoretical considerations of soliton growth by dynamical relaxation in order to convert the observational constraints on soliton mass into constraints on the ULDM-to-total-DM fraction $f$. In Sec.~\ref{s:com}, we comment on implications when ULDM is comprised of more than one species (``more than one $m$") and on a possible caveat related to soliton random walk. We conclude in Sec.~\ref{sec:conclusions}. In Appendix~\ref{app:modelsAB} we consider more realistic fitting procedures, including estimates of additional DM and baryonic mass components, to complement the more conservative analysis of the main text. Appendix~\ref{app:relax-mult-axions} outlines how the dynamical relaxation estimates can be generalized to a scenario with more than one species of ULDM. 

\section{ULDM versus rotation curve data}
\label{sec:backgrounds}

\subsection{soliton-halo relation: Summary of previous results}
Numerical simulations of ULDM  \cite{Schive:2014dra,Schive:2014hza} found an empirical relation, connecting the soliton to its host halo. While the soliton-halo relation was originally reported as a relation between the soliton mass and the host halo mass, 
\BarA and \BarB showed that the reported relation is precisely equivalent to a more physically tractable equality between the specific kinetic energy stored in the soliton and that in the host halo:
\begin{align}
\label{eq:SH-mass-rel}
\frac{K}{M}\Big|_{\rm sol}
\approx
\frac{K}{M}\Big|_{\rm halo}.
\end{align}
Here, $ K $ is the kinetic energy, and $ M $ is the total mass in each component (the soliton core on the lhs, and the host halo on the rhs) of the density profile. 

Phenomenologically, \refeq{SH-mass-rel} implies that the peak rotation velocity of test particles, induced by the soliton gravitational potential, should be close to the peak rotation velocity induced by the host halo. This shape information on the velocity curve makes the soliton-halo relation an easy observational target.

Theoretically, \refeq{SH-mass-rel} is suggestive of quasiequilibrium\footnote{By quasiequilibrium, we mean that the soliton may continue to grow after saturating the soliton-host halo relation, but at a parametrically reduced rate \cite{Eggemeier:2019jsu,Chen:2020cef}.} or approximate thermalization between ULDM particles in the halo and in the soliton structure. Such a behavior is consistent with the outcome of dynamical relaxation~\cite{Levkov:2018kau,Schwabe:2020eac} that is much more efficient in ULDM than in CDM models due to the formation of ULDM interference patterns or granules, acting as massive quasiparticles~\cite{Hui:2016ltb,Bar-Or:2018pxz}. 

In much of our analysis, we will use Eq.~(\ref{eq:SH-mass-rel}) as a benchmark for comparison of the soliton prediction of ULDM with observations. It is therefore important to emphasize that the soliton-halo relation as expressed by Eq.~(\ref{eq:SH-mass-rel}) is not without dispute. 
Reference~\cite{Mocz:2017wlg}, in particular, reported a different relation; however, it was shown in \BarA (see Sec.III.B there) that the soliton-halo relation of~\cite{Mocz:2017wlg} amounts  to precisely equating the entire total energy of the halo with that of the soliton, suggesting that the initial conditions adopted in Ref.~\cite{Mocz:2017wlg} were not realistic. More in general, additional numerical and analytical tests of Eq.~(\ref{eq:SH-mass-rel}) would be important\footnote{The tools developed in Ref.~\cite{Yavetz:2021pbc} may help in this direction.}: Our results strongly highlight this fact. We also note that for the purpose of deriving constraints on ULDM, the soliton-halo relation adopted in our benchmark analysis [Eq.~(\ref{eq:SH-mass-rel})] leads to conservative bounds -- with a conservative estimated uncertainty -- when compared to other scaling relations as reviewed in, e.g., Ref.~\cite{Chan:2021bja}.

In the remaining of this subsection we briefly review the derivation of Eq.~(\ref{eq:SH-mass-rel}) as given in \BarA and~\BarB. We also take this opportunity to explain in more detail the physical meaning of the different relation quoted in Ref.~\cite{Mocz:2017wlg} (clarifying what we believe is a critical caveat in that result). Readers who have followed the analysis in \BarA, or who are mainly interested in the observational consequences implied by Eq.~(\ref{eq:SH-mass-rel}), can skip to the following subsection without loss of information.

Originally, the simulation result of Refs.~\cite{Schive:2014dra,Schive:2014hza} was presented as a relation between the soliton mass and the host halo mass~\cite{Schive:2014dra}, which could be summarized by
\be M_{\rm sol}&\approx&1.4\times10^9\left(\frac{10^{-22}~{\rm eV}}{m}\right)\left(\frac{M_{\rm halo}}{10^{12}~{M}_\odot}\right)^{\frac{1}{3}}~{\rm M}_\odot.\;\;\;\ee
Reference~\cite{Schive:2014hza} noted another way by which the same result can be expressed; casting their result into natural units (see Sec.III.A of \BarA) it can be written as
\be\label{eq:SHorigin} M_{\rm sol}&=&\frac{\tilde\alpha}{Gm}\left(\frac{\left|E_{\rm halo}\right|}{M_{\rm halo}}\right)^{\frac{1}{2}},\;\;\;\;\tilde\alpha\approx4.2.\ee
Here, $G$ is the Newton constant, and $\tilde\alpha$ was an empirical ``fudge factor," extracted in~\cite{Schive:2014hza} by fitting to their simulation data.\footnote{To be precise, Ref.~\cite{Schive:2014hza} expressed their results in terms of the soliton ``core mass" $M_c$, defined as the mass enclosed by the soliton in the region where its density profile falls by a factor of 2 from its value at the center. Direct integration of the soliton profile gives the relation $M_{\rm sol}\approx4.2\,M_c$.}

This empirical picture was clarified to some extent in \BarA and~\BarB, as follows. 
The ULDM field under discussion is a massive free scalar field, that we denote by $\phi$. In the nonrelativistic limit (characteristic velocities much smaller than $c$), we can express $\phi$ in terms of the Schr\"odinger field $\psi$ as $\phi=1/\left(\sqrt{2}m\right)e^{-imt}\psi({\bf x},t)+c.c.$,\footnote{Note that we follow \BarA and \BarB notation so $\psi$ has mass dimension two, with $|\psi|^2$ being the mass density instead of number density.} where in the relevant limit $|\nabla\psi|\ll m|\psi|$, $|\dot\psi|\ll m|\psi|$.
The soliton is a spherically symmetric self-gravitating ground-state solution of the Schr\"odinger-Poisson equations of motion of $\psi$ (nonrelativistic limit of the full Einstein-Klein Gordon equations of motion of $\phi$); this solution can be straightforwardly computed numerically. The total mass, total energy, and total kinetic energy can be expressed as functionals of the field:
\be M_{\rm sol}&=&4\pi\int dxx^2\left|\psi(x)\right|^2,\\
E_{\rm sol}&=&4\pi\int dxx^2\left(\frac{1}{2m^2}\left|\nabla\psi(x)\right|^2+\frac{1}{2}\Phi(x)\left|\psi(x)\right|^2\right),\;\;\;\;\\
 K_{\rm sol}&=&\frac{4\pi}{2m^2}\int dxx^2\left|\nabla\psi(x)\right|^2.\ee
By direct integration of these functionals for the soliton solution, \BarA and~\BarB showed that the solution is virial, that is $K_{\rm sol}=-E_{\rm sol}$; and, moreover, it satisfies the relation
\be\label{eq:relation} M_{\rm sol}&\approx&\frac{4.3}{Gm}\left(\frac{\left|K_{\rm sol}\right|}{M_{\rm sol}}\right)^{\frac{1}{2}}.\ee
In this expression, both the lhs and the rhs apply to a self-gravitating ``stand-alone" soliton solution. 

Now, compare Eq.~(\ref{eq:relation}) to the numerical simulations of Refs.~\cite{Schive:2014dra,Schive:2014hza}, as summarized by Eq.~(\ref{eq:SHorigin}). There, the lhs of the equation is again just the soliton mass, while the rhs expresses the result of the simulation for the incoherent {\it large-scale host halo}, at the center of which the soliton is detected. Given that the central soliton observed in simulations is very well described by the self-gravitating solution; and assuming that the host halo is approximately virialized as well, satisfying $E_{\rm halo}\approx-K_{\rm halo}$, we can conclude that the empirical result of Refs.~\cite{Schive:2014dra,Schive:2014hza} is contained by equating the rhs of Eq.~(\ref{eq:relation}) (referring to the soliton) with the rhs of Eq.~(\ref{eq:SHorigin}) (referring to the host halo). This is the content of Eq.~(\ref{eq:SH-mass-rel}).

Let us now apply a similar exercise to the results claimed in Ref.~\cite{Mocz:2017wlg}. Again, we follow the discussion in \BarA (Sec.III.B there). The ``soliton-halo relation" claimed by Ref.~\cite{Mocz:2017wlg} was (in natural units)
\be\label{eq:Mocz} GmM_{\rm sol}&\approx&2.6\left(Gm\left|E_{\rm halo}\right|\right)^{\frac{1}{3}}.\ee
The numerical factor of $\approx2.6$ was derived empirically by the authors of Ref.~\cite{Mocz:2017wlg}, fitting their simulation results. 

Alas, a direct integration of the soliton field functionals, done in \BarA, reveals that a self-gravitating soliton satisfies
\be\label{eq:Mocz0} GmM_{\rm sol}&\approx&2.64\left(Gm\left|E_{\rm sol}\right|\right)^{\frac{1}{3}}.\ee
Again, the left-hand sides of both Eqs.~(\ref{eq:Mocz}) and~(\ref{eq:Mocz0}), and the rhs of~(\ref{eq:Mocz0}), refer to the soliton, while the rhs of Eq.~(\ref{eq:Mocz}) refers to the large-scale host halo as found in the simulations of~\cite{Mocz:2017wlg}. Equating the right-hand sides of Eqs.~(\ref{eq:Mocz}) and~(\ref{eq:Mocz0}), we can conclude that the entire ``soliton-halo relation" of Ref.~\cite{Mocz:2017wlg} can be precisely summarized by noting that this study produced halos with a total energy that was completely dominated by their central solitons: $E_{\rm halo}\approx E_{\rm sol}$. This relation cannot be expected to hold for real massive cosmological halos (satisfying $M_{\rm halo}\gg M_{\rm sol}$). Instead, we suspect that the scaling claimed by Ref.~\cite{Mocz:2017wlg} was an artifact of the initial conditions chosen for their numerical experiment, which {\it was not} the result of cosmological initial conditions for the ULDM. 
More discussion of the details and impact of these initial conditions can be found in \BarA.\footnote{We must comment here that numerical experiments in Ref.~\cite{Schive:2014hza} also employed toy simulations with noncosmological initial conditions. Importantly, however, these toy simulations: (i) were shown to agree with the scaling observed in actual cosmological simulations in Ref.~\cite{Schive:2014dra}, and (ii) employed initial conditions which were essentially different to those in Ref.~\cite{Mocz:2017wlg}.} 

We conclude that the soliton-halo relation claimed in  Ref.~\cite{Mocz:2017wlg} simply says that in these simulations, the entire total energy of the ``host halo" was dominated by a single soliton, a situation that is unlikely to describe realistic cosmological ULDM halos. In contrast, the relation obtained in Refs.~\cite{Schive:2014dra,Schive:2014hza}, summarized by \BarA and \BarB in terms of Eq.~(\ref{eq:SH-mass-rel}), {\it could} be physical and was indeed discovered by Ref.~\cite{Schive:2014dra} in simulations utilizing cosmological initial conditions. We believe that the theoretical perspective we reviewed here did not receive full attention in some assessments, such as Ref.~\cite{Chan:2021bja}. Having clarified our perspective on this matter, in the rest of the paper we focus on Eq.~(\ref{eq:SH-mass-rel}) as a physically motivated benchmark for our results.

\subsection{Looking for solitons in SPARC}\label{ssec:data}
We use rotation curve data from the SPARC database \cite{Lelli:2016zqa} to look for the imprint of solitons. The database consists primarily of observationally inferred rotation curve data, along with model results aiming to separate the contribution of baryons (stellar disk and bulge, as well as gas), for 175 nearby galaxies. 

In our main and most conservative pass on the data, we ignore the modeling attempts to identify the baryonic contribution to the rotation curve. Instead, to constrain the allowed $ M_{\rm sol} $ in a given galaxy, we perform a ``one-sided'' test, where a soliton contribution is excluded if it alone overshoots some portion of the rotation curve data to some specified significance. This approach is equivalent to modeling a soliton together with an arbitrary background profile, where the background profile can be adjusted to fit any velocity bin that the soliton-induced velocity undershoots. The only assumption we make for the (otherwise unspecified) background component is that it gives a positive contribution to the rotation velocity. 

The radial mass profile due to the soliton, $M(r)$, is given by
\begin{align}
\label{eq:model-A-M}
M(r', M_{\rm sol}, m)
& =
\int_0^{r'} \rho_{\rm sol}(r; M_{\rm sol}, m) \; 4\pi r^2 dr,
\end{align}
where the soliton density profile is given approximately by~\cite{Schive:2014hza}
\begin{align}
\label{eq:sol-profile}
\rho_{\rm sol}(r)
& \approx
\frac{    \rho_{\rm sol}(0) }{(1 + 0.091 (r/r_c)^2)^8},
\\
\rho_{\rm sol}(0)
& \approx
0.083 \; \left ( \frac{M_{\rm sol}}{M_\odot} \right )
\left ( \frac{r_c}{\mathrm{kpc}} \right )^{-3}
\; \frac{M_\odot}{\mathrm{kpc}^3},\no 
\end{align}
for a soliton total mass $M_{\rm sol}$. The characteristic radius $r_c(M_{\rm sol}, m)$ is given by
\begin{align}
\label{eq:core-radius}
r_{c}
& \approx
2.28
\left ( \frac{M_{\rm sol}}{10^{11} M_{\odot} } \right )^{-1}
\left ( \frac{m}{10^{-22} \; \mathrm{eV}} \right )^{-2}
\; \mathrm{pc}.
\end{align}
The soliton profile is then controlled by two parameters, $M_{\rm sol}$ and $m$. In our analysis we scan a fixed grid in $m$, determining the limit on $M_{\rm sol}$ for each value of $m$.

Note that we use the self-gravitating soliton profile, without including the distortion of the profile due to the presence of the non-ULDM background density. The effect of the background density was studied in detail in Ref.~\BarB. The general results of that analysis indicate that the self-gravitating soliton profile is a good approximation to the actual profile, as long as the background mass component contained within the soliton core radius is smaller than the total soliton mass. While this assumption can be violated when the soliton mass is small, it is valid in the region of the $M_{\rm sol},m$ parameter space that saturates our bound.

The results of our analysis are shown in \reffig{final-result-intro}. 
The parameter region over which the data are most sensitive to the soliton-halo relation is $m\sim10^{-21}-10^{-22}$~eV. As discussed in \BarA and Ref.~\BarB, we find no convincing hint for the soliton bump in any well-resolved, DM-dominated rotation curve. We therefore present exclusion limits, extending the discussion in \BarA. 

For very large and very small values of $ m $, we can understand the scaling of the exclusion curves in \reffig{final-result-intro} analytically. This is highlighted in \reffig{final-result-intro} as dashed lines for one sample galaxy. At low $ m $, the constraints are dominated by the largest radius data bin $ r_{\rm f} $, which falls inside the soliton core.  
The largest radius bin therefore constrains $ M_{\rm sol}(r_{\rm f}) < r_{\rm f}V^2_{\rm obs}(r_{\rm f})/G$~=~const., where $V_{\rm obs}$ is the observed rotation velocity. 
In this regime, the enclosed soliton mass is $ M_{\rm sol}(r_{\rm f})\approx (r_{\rm f}/r_c)^3M_{\rm sol} \propto M_{\rm sol}^4 m^6$ [using \refeq{core-radius}]. 
In the plot, we show the scaled ratio, $M_{\rm sol}/M_{\rm SH}$, where $M_{\rm SH}\propto m^{-1}M_{\rm halo}^{{1}/{3}}$~\cite{Schive:2014dra,Schive:2014hza}.  
 Therefore, the constraint on the ratio $M_{\rm sol}/M_{\rm SH}$ in the low-$m$ region in \reffig{final-result-intro} follows $ M_{\rm sol}/M_{\rm SH}\propto (m^{-{6}/{4}})/(m^{-1})\propto m^{-1/2} $.
At large $m$, the constraint is dominated by the innermost data bin $ r_{\rm i} $, and the soliton potential is approximately that of a point mass. The data then constrain the total soliton mass $ M_{\rm sol} $, so the constraint on $ M_{\rm sol}/M_{\rm SH} \propto m $. 

In the remaining part of this section, we discuss a number of additional points related to the constraints in \reffig{final-result-intro}.

\textbf{Plateau at small $m$: lack of constraining power for high-surface-brightness galaxies.}
It is interesting to note that, in the small-$m$ region $m\lesssim10^{-24}$~eV, the data (including potential sensitivity from many galaxies) are compatible with a soliton saturating the soliton-halo relation. Observationally, this reflects the fact that many galaxies in the SPARC database display rotation curves that scale linearly with radius, $V_{\rm obs}\propto r$, consistent with the total density profiles of these galaxies forming large-radius cores. The shallow slope of these rotation curves suggest low-density, large-radius cores; to attribute such cores to ULDM, one would be forced to require $m\lesssim10^{-24}$~eV. Such light ULDM is in strong contradiction with cosmological Ly-$\alpha$ data, unless the ULDM makes up just a small fraction of the total DM, $f\lesssim0.2$. If that was the case, it is unclear how the main 80\% of the DM disappears from these galaxies.

\begin{figure}[th]
	\centering
        \includegraphics[width=.495\textwidth]{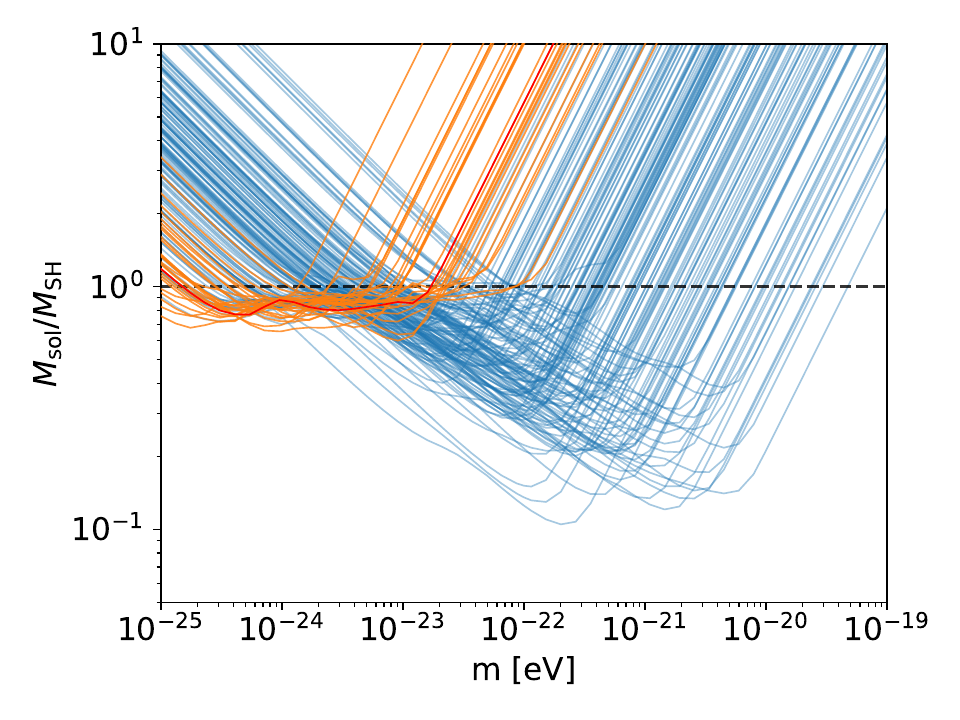}
        \includegraphics[width=.495\textwidth]{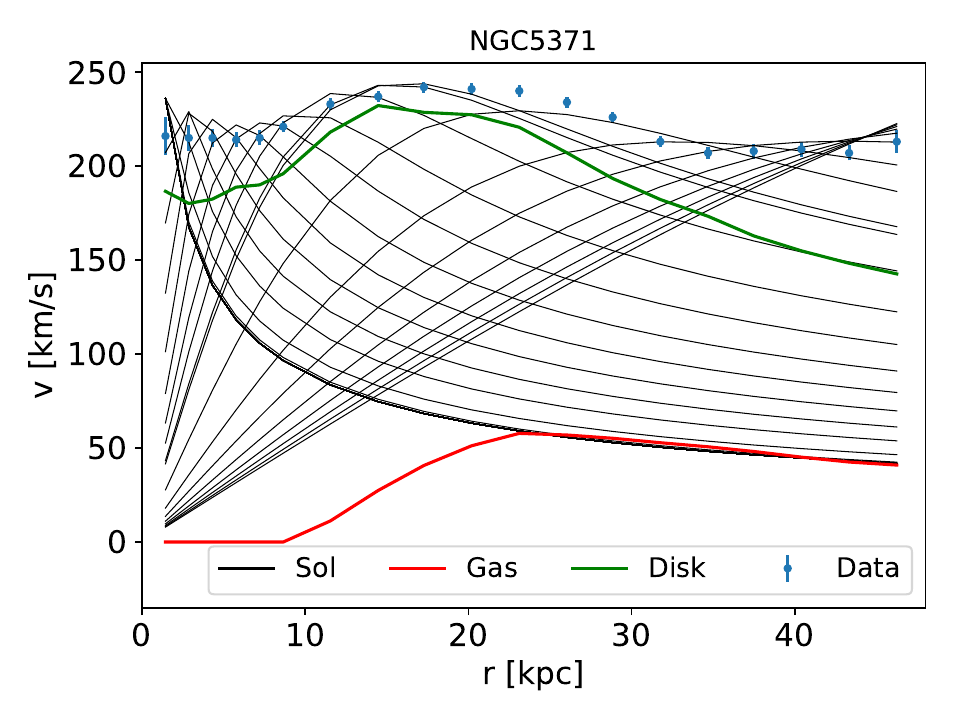}
	\caption{\textbf{Top panel:} similar to \reffig{final-result-intro} but highlighting (in orange) bounds arising from galaxies with particularly weak constraints at $m>10^{-23}\;\mathrm{eV}$. The bound from NGC5371, which is a typical system for this set of rotation curves, is colored red (embedded in the orange lines). \textbf{Bottom panel:} detailed rotation curve of NGC5371 (see the text), including the estimated baryonic gas (red) and disk (green) components. Soliton components (black) saturating the bound in the top panel are also shown for values of $m$ in the range $10^{-25}\;\mathrm{eV}$ to $10^{-19}\;\mathrm{eV}$.}
        \label{fig:check-flat}
\end{figure}

To clarify this point further, in the top panel in Fig.~\ref{fig:check-flat}, we highlight the bounds corresponding to  galaxies that exhibit a flat segment at small $m$. 
Referring to SPARC data, we find most of these galaxies are high-surface-brightness galaxies with a large baryonic component. As a representative example, we focus on NGC5371; in the bottom panel in Fig.~\ref{fig:check-flat}, we plot the observed rotation curve (blue markers), along with an estimated contribution of each baryonic component stellar disk (green) and gas (orange). For the disk, we assume $\Upsilon_{\rm disk} = 0.6$. We then superimpose the contribution of a soliton component with different values of $m$ and normalization chosen to saturate the bound depicted for this system in the top panel (thin red line, embedded in the family of orange lines in the top panel). This inspection makes clear that the weak constraint on $m$ arising from this galaxy is a consequence of our conservative baseline analysis, which does not attempt to subtract any model of the baryonic components, but rather just requires the soliton not to overshoot the observed velocity. 
      
%
\begin{figure}[th]
	\centering
        \includegraphics[width=.495\textwidth]{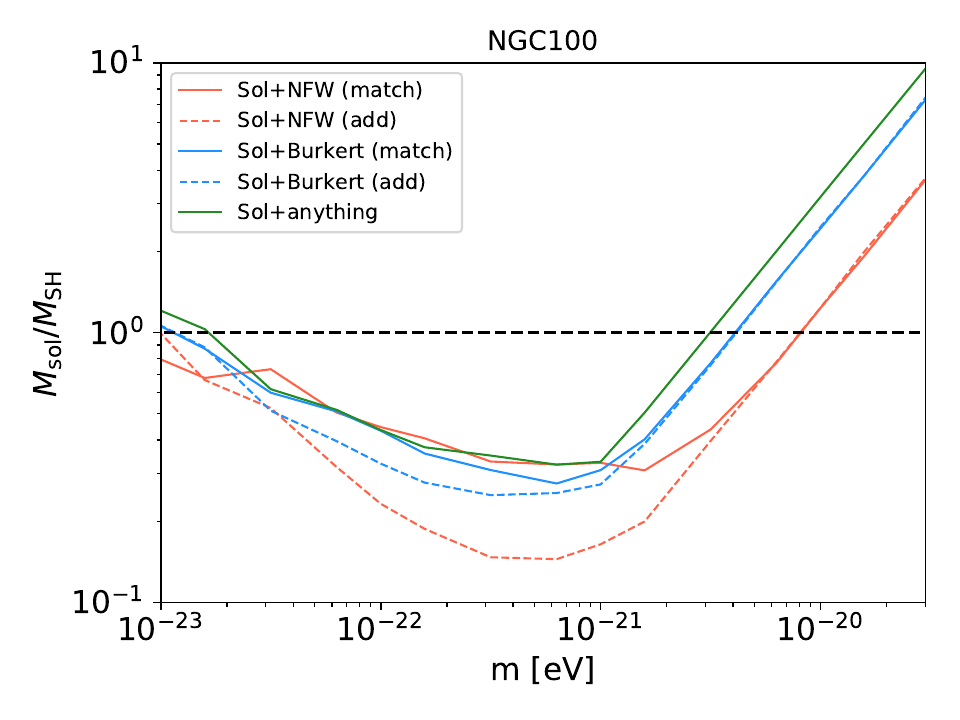}
	\caption{95\%~C.L. limit on $M_{\rm sol}/M_{\rm SH}$ versus $m$, obtained with different modeling of the background density profile for a sample galaxy (NGC100). The horizontal dashed line marks the soliton-halo relation. The green solid line shows the limit obtained with an arbitrary background profile: This is the procedure we refer to in the main text. In addition, we also show results where the background DM is fitted with NFW (red) or Burkert (blue) profiles, matching the soliton and background halo components (solid) as well as simply adding the components on top of each other (dashed). More details can be found in Appendix~\ref{app:modelsAB}. 
	}  
	\label{fig:compare-five-models}
\end{figure}
\textbf{Realistic background density.}
The analysis leading to \reffig{final-result-intro} was conservative, in that we considered the impact of the soliton only when it overshoots the rotation velocity data, allowing an unspecified background density profile to fit underpredicted velocity bins. In Appendix~\ref{app:modelsAB} we study how the limits change when considering more realistic background profiles. The exercise there involves a statistical fit of the velocity profile, deriving the constraints on ULDM from a log-likelihood ratio. In addition to the soliton component, we include the following ingredients. (i) We add the baryonic contribution to the velocity curve, using the gas, disk, and bulge models from SPARC~\cite{Lelli:2016zqa} and allowing the mass-to-light ratios of the disk and the bulge to vary freely in the fit; (ii) we consider two models for the DM contribution, in addition to the soliton: A Navarro-Frenk-White (NFW) profile~\cite{Navarro:1996gj}, and a cored Burkert profile~\cite{Burkert:1995yz}. These are matched to the soliton feature in different ways.

We leave the details of the fitting analysis to Appendix~\ref{app:modelsAB}. The results are shown in \reffig{compare-five-models}. In terms of the limit on $M_{\rm sol}/M_{\rm SH}$ or on $m$, the consideration of more realistic background profiles strengthens the limit by up to a factor of 2  in the large $m$ region. 
%
%

\begin{figure}[th]
	\centering
	\includegraphics[width=.495\textwidth]{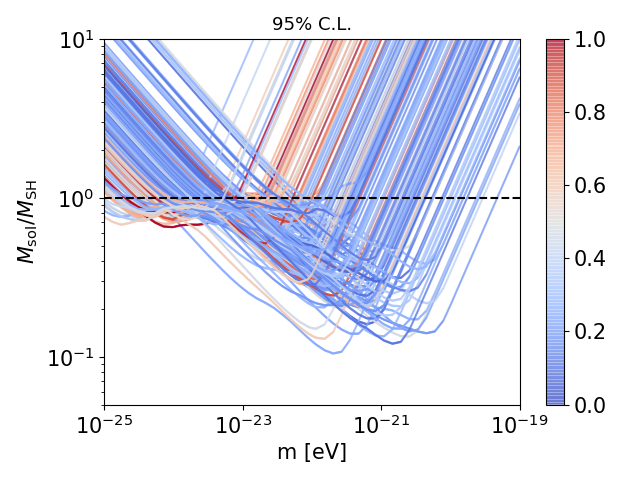}
        \includegraphics[width=.495\textwidth]{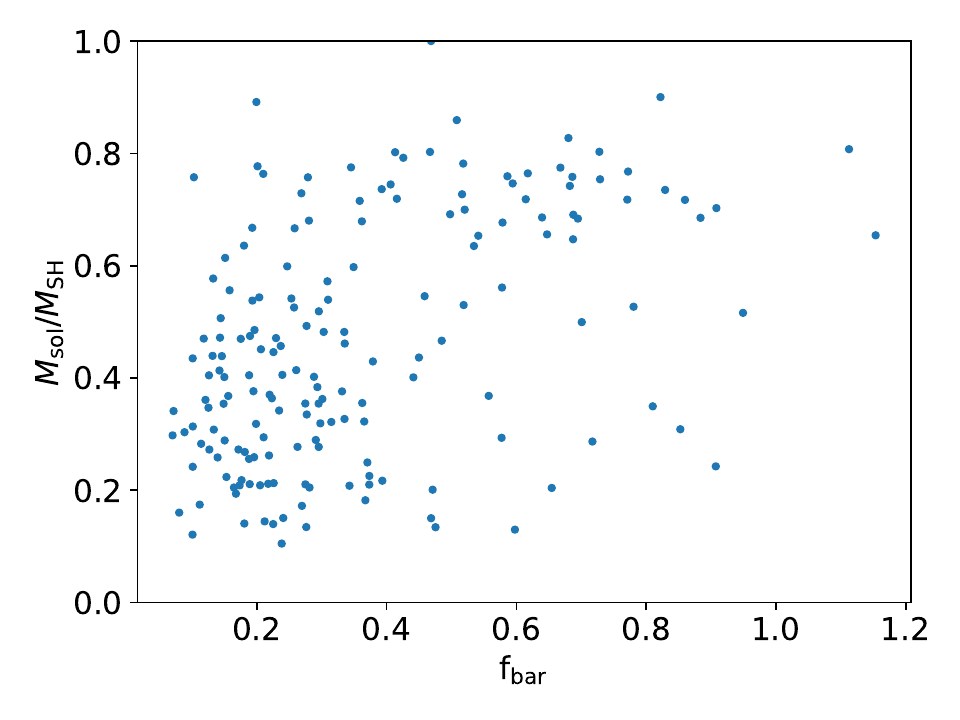}
	\caption{\textbf{Top panel:} same as \reffig{final-result-intro}, but color-coding the baryonic fraction $\frac{M_{\rm bar}}{M_{\rm tot}}\Big|_{r_{\rm peak}}$ [see the text around Eq.~(\ref{eq:bar2dm})] for each galaxy. \textbf{Bottom panel:} scatter plot including all galaxies in the sample, showing the baryonic fraction defined in Eq.~(\ref{eq:bar2dm}) on the $x$ axis, and the tightest constraint on $M_{\rm sol}/M_{\rm SH}$ derived for each galaxy on the $y$ axis (namely, the $M_{\rm sol}/M_{\rm SH}$ quoted for the value of $m$ at which the bound is strongest).}
	\label{fig:f-cut}
\end{figure}

\textbf{Impact of baryons.}
In the top panel in \reffig{f-cut}, we repeat \reffig{final-result-intro}, color coding the limit from each galaxy according to the importance of the baryonic contribution in the rotation curve. The baryonic contribution is estimated via 
\be\label{eq:bar2dm}
\frac{M_{\rm bar}}{M_{\rm tot}}\Big|_{r_{\rm peak}} &=& \frac{V_{\rm bar}^2}{V_{\rm obs}^2}\Big|_{r_{\rm peak}} ,
\ee
where $ r_{\rm peak} $ is the bin with maximal rotation velocity.  For the purpose of this estimate we fix the mass-to-light ratios as $ \Upsilon_{\rm disk} = 0.5 $ and $ \Upsilon_{\rm bulge} = 0.5 $ (see Appendix~\ref{app:modelsAB} for details). 
We see that (i) the strongest constraints derive mostly from DM-dominated galaxies, with $\frac{M_{\rm bar}}{M_{\rm tot}}\Big|_{r_{\rm peak}} < 0.5$, and (ii)~dropping galaxies with a high baryonic fraction from the analysis would not affect the results. In the bottom panel of \reffig{f-cut} we further explore the impact of the baryonic fraction by means of a scatter plot, showing that the strongest constraints, again, arise from DM-dominated systems.

  \textbf{Role of host halo mass.}
We now inspect the role played by the mass of the host halo in a given galaxy in the soliton bound derived for that galaxy. To this end, we define a proxy for the virial mass of the host halo as
\begin{align}
  \label{eq:halo-virial-mass-proxy}
  M_{\rm halo}
  & =
    \mathrm{max}_{r} \left (
    \frac{V^2_{\rm obs}(r) r}{G}
    \right ),
\end{align}
where $V_{\rm obs}$ is the observed rotation velocity, and the maximization is carried with respect to all radius bins. 

\begin{figure}[th]
  \centering
	\includegraphics[width=.49\textwidth]{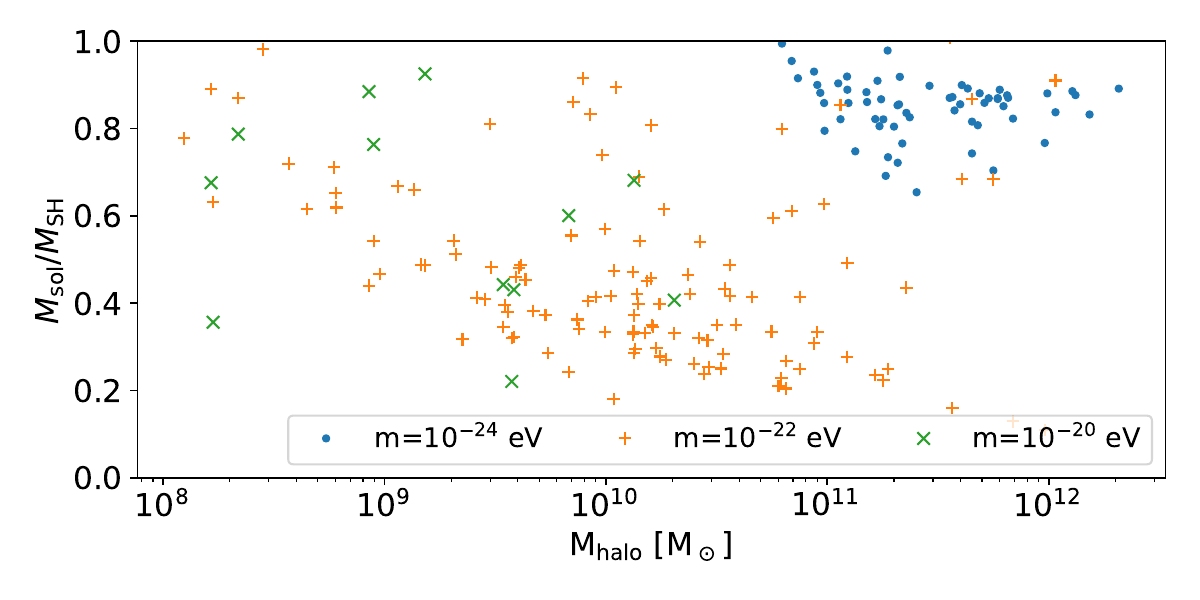}  
	\caption{Scatter plot of the $M_{\rm sol}/M_{\rm SH}$ bound versus halo mass (inferred from the rotation curve; see text), for all galaxies in the sample, specified at three reference values of $m$: $10^{-24}\;\mathrm{eV}$ (blue dot), $10^{-22}\;\mathrm{eV}$ (orange $+$), $10^{-20}\;\mathrm{eV}$ (green $\times$).} 
	\label{fig:scattering-Mhalo}
\end{figure}

In 
Fig.~\ref{fig:scattering-Mhalo} we show a scatter plot of the $M_{\rm sol}/M_{\rm SH}$ bound versus $M_{\rm halo}$, obtained for three representative values of $m$. For clarity, we truncate the $y$ axis at $M_{\rm sol}/M_{\rm SH}\leq1$; namely, we show only those systems which place an informative limit on ULDM. We can see that more massive galaxies place the most important constraints for small values of $m$ (blue dots corresponding to $m=10^{-24}$~eV), while lower mass galaxies are most important at larger $m$ (orange $+$ and green $\times$ corresponding to $m=10^{-22}$~eV and $m=10^{-20}$~eV, respectively). The main reason for this is simply the data coverage of different types of galaxies in the sample: The data for massive galaxies often extend out to many kiloparsecs, allowing one to probe the slow-rising soliton profiles of low-$m$ ULDM, but is not well resolved at small $r\ll1$~kpc and, thus, cannot constrain the abrupt feature induced by large-$m$ ULDM. Low mass galaxies have the opposite trend.  

%

\begin{figure}[th]
	\centering
	\includegraphics[width=.495\textwidth]{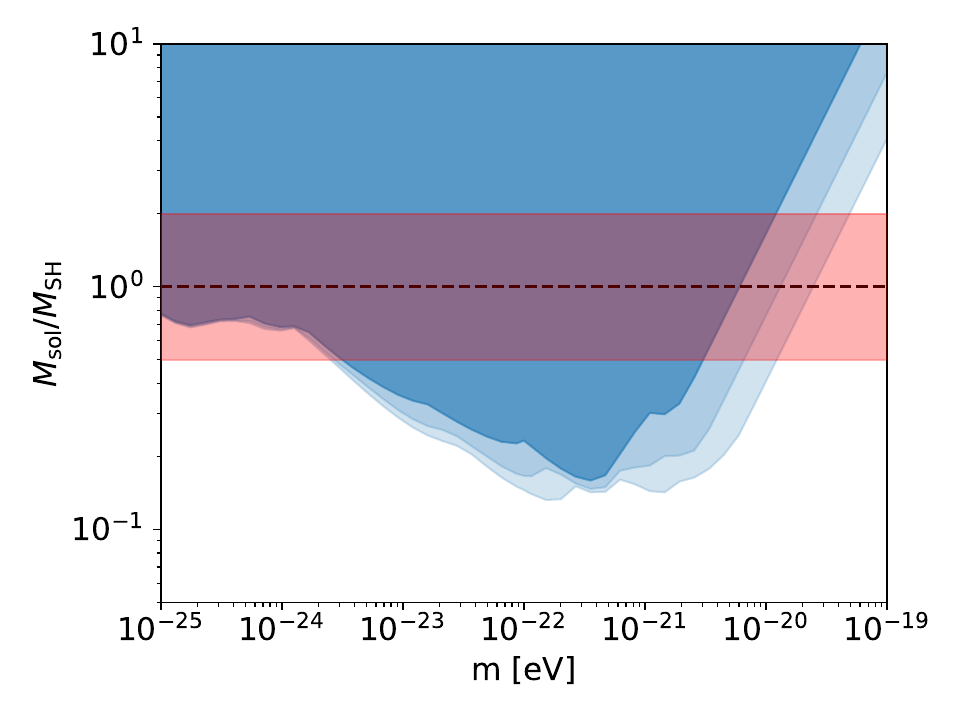}
	\caption{
		Combined $3\,\sigma, 5\,\sigma, 10\,\sigma$ constraints on the soliton-halo relation. The axes are the same as Fig.~\ref{fig:final-result-intro}.}
	\label{fig:final-plot}
\end{figure}

\textbf{Statistical significance.}
Figure~\ref{fig:final-plot} compares the $ 3\,\sigma, 5\,\sigma $, and $ 10\,\sigma$ constraints obtained by combining the data from all of the SPARC galaxies. At large $m\gtrsim10^{-21}$~eV, the difference between the 3~$\sigma$ and 10~$\sigma$ excluded regions, in terms of $M_{\rm sol}/M_{\rm SH}$ or $m$, is roughly a factor of 2.

\section{Constraining the ULDM fraction}
\label{sec:interpr-constr-solit}
The constraints we derived in Sec.~\ref{ssec:data} on $ M_{\rm sol}/M_{\rm SH} $ versus $ m $ were purely observational: We simply looked in the data for the imprint of the soliton core, and constrained its possible amplitude. The role of the theoretical quantity $M_{\rm SH}$ in that exercise was simply to provide a convenient reference point, so that  results from different galaxies could be analyzed in conjunction. In the current section, our goal is to turn these observational limits into constraints on the fraction $f$ of the total DM, which could be supplied by ULDM. To do this, we need to understand under what conditions the soliton-halo relation, $M_{\rm sol}\approx M_{\rm SH}$ is expected to hold in reality.

An irreducible channel for the formation of soliton cores is via gravitational dynamical relaxation~\cite{Hui:2016ltb,Levkov:2018kau,Bar-Or:2018pxz} acting on an initially incoherent ensemble of ULDM waves.\footnote{Dynamical relaxation starting from a stochastic initial state is not necessarily the only channel to form solitons. In principle, a coherent soliton core could exist in halo centers from the early structure formation stage.} 
The dynamical relaxation time for ULDM in a system with ULDM density $\rho$ and one-dimensional velocity dispersion $\sigma$ is~\cite{Hui:2016ltb,Levkov:2018kau,Bar-Or:2018pxz}
\begin{align}
 \label{eq:relax-time-ULDM}
 \tau
 & =
 \frac{b \sqrt{2}}{12 \pi^3} \frac{ m^3 \sigma^6}{ G^2 \rho^2 \ln(m \sigma R)} \\
& \approx 
10~\mathrm{Myr} 
\left (\frac{m}{10^{-22} \; \mathrm{eV}}\right )^3
\left ( \frac{\sigma}{50~\frac{\mathrm{km}}{\mathrm{s}}} \right )^6
\left ( \frac{0.1 \frac{M_\odot}{\mathrm{pc}^3}}{\rho} \right )^{2}\left(\frac{3}{\ln\Lambda}\right) .\no
 \end{align}
The numerical factor $ b \approx 0.7 $ is calibrated by numerical simulations \cite{Levkov:2018kau} (see also \cite{Eggemeier:2019jsu,Chen:2020cef,Schwabe:2020eac}). We estimate the Coulomb log as $\ln\Lambda=\ln\left(m\sigma R\right)$, where $R$ is the characteristic radius of the system. Note that Eq.~\eqref{eq:relax-time-ULDM} is expected to become inaccurate for $\ln\Lambda\lesssim1$. 

Equation~(\ref{eq:relax-time-ULDM}) shows that over wide regions in the density profile of typical galaxies (specifically, typical SPARC galaxies referred to later on in this work), $\tau$ can become much shorter than the age of the galaxy.

The relaxation time becomes longer if ULDM comprises only a fraction $f<1$ of the total density $ \rho $; in that case, we should replace $\rho\to f\rho$ in Eq.~\eqref{eq:relax-time-ULDM}~\cite{Blum:2021oxj} (see also Appendix~\ref{app:relax-mult-axions}
).

If the relaxation time is much shorter than the age of a galaxy, then we expect that a soliton should form. Once the soliton specific kinetic energy saturates the value corresponding to the soliton-halo relation, Eq.~\eqref{eq:SH-mass-rel} (that is, once $M_{\rm sol}$ grows to saturate $M_{\rm SH}$), the soliton growth by dynamical relaxation becomes quenched and slows down considerably. This scenario, which is understood theoretically, is consistent with the results of numerical simulations \cite{Schive:2014dra,Schive:2014hza,Levkov:2018kau,Eggemeier:2019jsu,Chen:2020cef}.

On the other hand, a relaxation time longer than the age of a galaxy may mean that a soliton could not have formed in the galaxy. In such a system, we do not translate the observational constraints on $M_{\rm sol}$ to a constraint on total ULDM fraction $f$.  

Following Ref.~\cite{Blum:2021oxj} we suggest a concrete, approximate criterion, to see if a given galaxy should be expected to have formed a soliton of mass $M_{\rm sol}$ by dynamical relaxation. To this end, we define two characteristic radii:
\begin{itemize}
\item $r_{\rm supply}$: If solitons grow by accreting mass from an initially stochastic halo, then to assemble a soliton of mass $M_{\rm sol}$, field needs to be accreted from a radius that is at least as large as $ r_{\rm supply} $, defined by $ \int_0^{r_{\rm supply}} 4\pi r^2\rho_X(r)dr = M_{\rm sol} $, where $ \rho_X $ is the initial ULDM halo density profile. If ULDM makes up only a fraction of the total DM density, then only the ULDM part should be included in $\rho_X$. In particular, if the ULDM fraction $f$ is decreased, then $r_{\rm supply}$ must {\it increase}, to compensate for the overall smaller ULDM density by drawing mass from larger distances. 

To make an analytic estimate, if $\rho_X$ follows an NFW profile, then for $ r\ll r_s $ we have $ \rho_X\propto f/r$ and $ M_{\rm sol}\propto fr_{\rm supply}^2 $. Since $ M_{\rm SH}\propto 1/m $, we find $ r_{\rm supply}\propto (mf)^{-1/2} $. If, for a very massive soliton, the process extends out to the region $ r\sim r_s $, where $ \rho_X \propto f/r^2 $, a similar consideration gives $ r_{\rm supply}\propto (mf)^{-1} $.
\item $r_{\rm relax}$: The process of soliton growth should be efficient only within a region of the halo for which the dynamical relaxation time is shorter than the age of the system. Defining the boundary of that region by $ r_{\rm relax} $, we have: $ \tau_{\rm relax}(r_{\rm relax}) = t_{\rm gal} $, with $t_{\rm gal}$ the age of the galaxy. As discussed above, if ULDM makes up only a fraction $f<1$ of the total DM density, then $\tau$ in Eq.~\eqref{eq:relax-time-ULDM} is increased as $\tau\to \tau/f^2$. Thus, making $f$ smaller has the effect of pushing $r_{\rm relax}$ further {\it in} to a smaller radius in the halo, to compensate for the smaller $f$ by a larger density (for simplicity, in this argument we assume a roughly constant velocity dispersion $\sigma$; in practice, we use a prescription to estimate $\sigma(r)$ from the observed velocity data, to be explained shortly below).

  To make an analytic estimate, assuming $ \sigma \approx$~const. one finds $ r_{\rm relax} \propto f/(m^3\sigma^6)^{1/2} $ for $ r\ll r_s $, and $ r_{\rm relax} \propto f^{1/2}/(m^3\sigma^6)^{1/4} $ at $ r\sim r_s $.
  In the numerical computation, we take the initial ULDM density to follow an NFW profile, and estimate $\sigma$ using Jeans modeling, discussed below.  
\end{itemize}
A rough criterion for the formation of a soliton with mass $M_{\rm sol}$ is 
\be\label{eq:SHcrit}
 r_{\rm relax}(M_{\rm sol},m,f) > r_{\rm supply}(M_{\rm sol},m,f).
 \ee
We take this as a condition for the applicability of the soliton-host halo relation. 
%

To estimate the local velocity dispersion in the relaxation time in Eq.~(\ref{eq:relax-time-ULDM}), we solve the Jeans equation for self-gravitating NFW halo, assuming isotropic velocity dispersion~\cite{binney2011galactic}
\be
\sigma^2(r) = \frac{G}{\rho(r)}\int\limits_r^{\infty} \frac{\rho(r^{\prime})M(r^{\prime})}{r^{\prime 2}}dr^{\prime} \; ,
\ee
where $ M(r) = \int_0^rd^3r^{\prime}\rho(r^{\prime}) $ is the enclosed mass. For NFW, this integral has an analytic solution. To speed up the numerical analysis, we use an approximate form for $ \sigma $, $\sigma(r)/V_{\rm circ}(r) \approx 0.55 + 0.2\; \mathrm{exp} \left ( - {r}/{2r_s} \right ) + 0.2\; \mathrm{exp} \left ( - {2r}/{r_s} \right ) + 0.6\; \mathrm{exp}\left (- {8r}/{r_s} \right ),       $
where $r_s$ is the transition scale in NFW as defined in Eq.~(\ref{eq:NFW-rho1}) and $V_{\rm circ}(r)$ is the circular velocity. The approximation differs from the exact solution by less than 2\% in the range of $0.05 \,r_s < r < 10\, r_s$.

\begin{figure}[ht]
	\includegraphics[width=.495\textwidth]{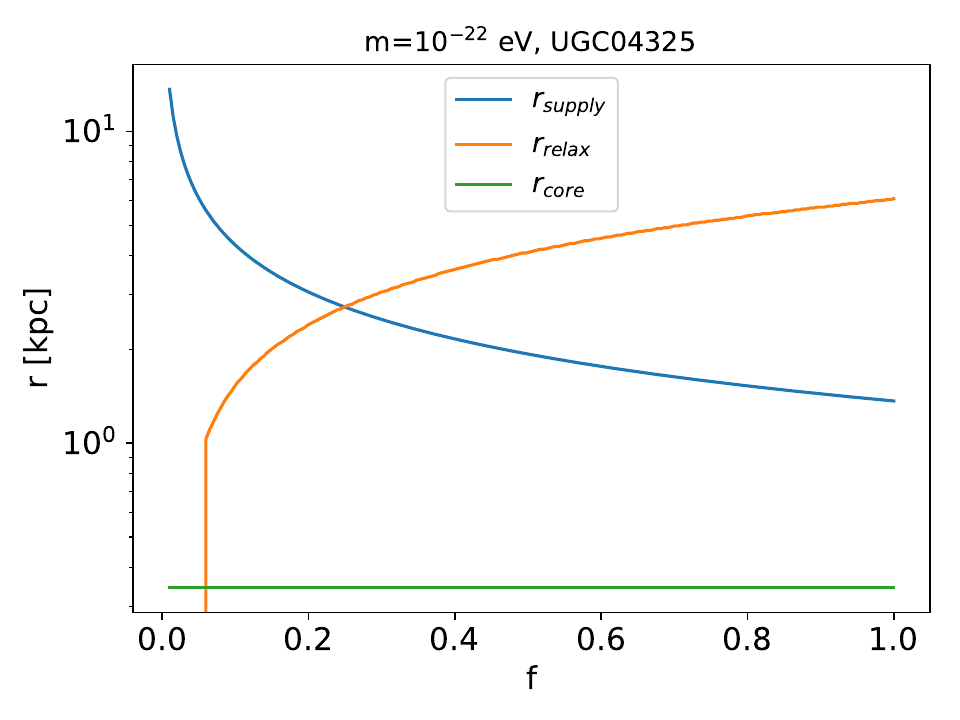}
	\caption{Radial scales entering the dynamical relaxation criterion, Eq.~\eqref{eq:SHcrit}, versus the cosmological ULDM fraction $ f $. In this example we consider the galaxy UGC4325, and set $m = 10^{-22}$~eV. }
	\label{fig:r-scaling-f}
\end{figure}

In \reffig{r-scaling-f} we show the different scales as functions of the ULDM fraction $ f $, for one sample galaxy, setting $m = 10^{-22}$~eV. We also show the core radius $r_c$ of a soliton that satisfies $M_{\rm sol}=M_{\rm SH}$ for this system. In this galaxy, according to the criterion Eq.~\eqref{eq:SHcrit}, ULDM with the prescribed value of $m$ can be expected to form a soliton saturating the soliton-halo relation only for $ f\gtrsim 0.3$. For smaller values of $f$, the soliton-host halo relation may break down as dynamical relaxation becomes inefficient.

\begin{figure*}[t]
  \includegraphics[width=.45\textwidth]{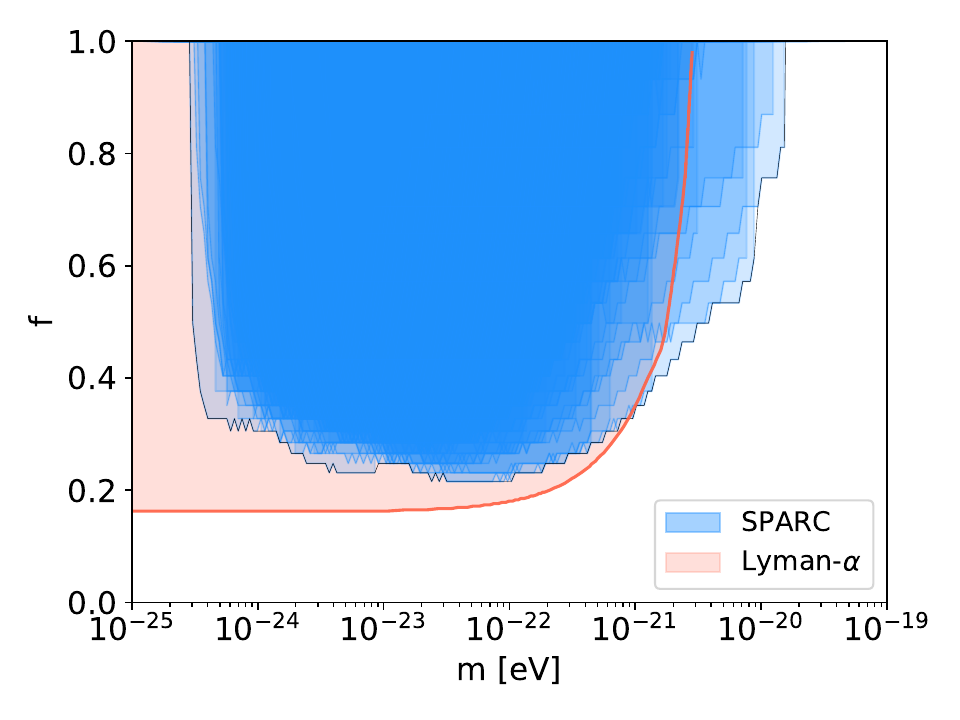}
  \includegraphics[width=.45\textwidth]{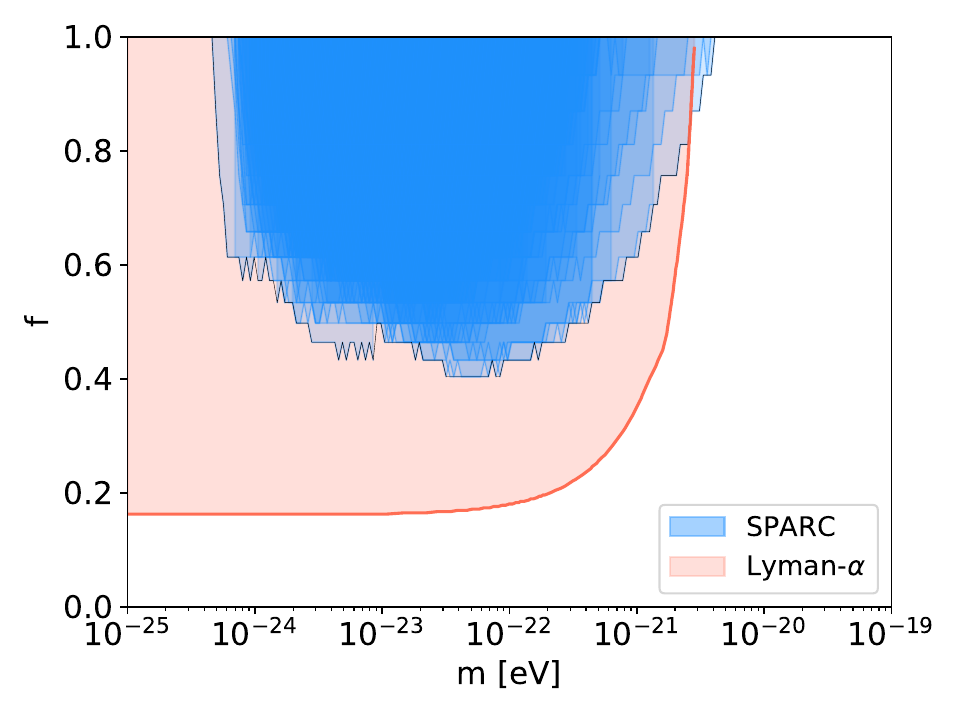}
  \includegraphics[width=.45\textwidth]{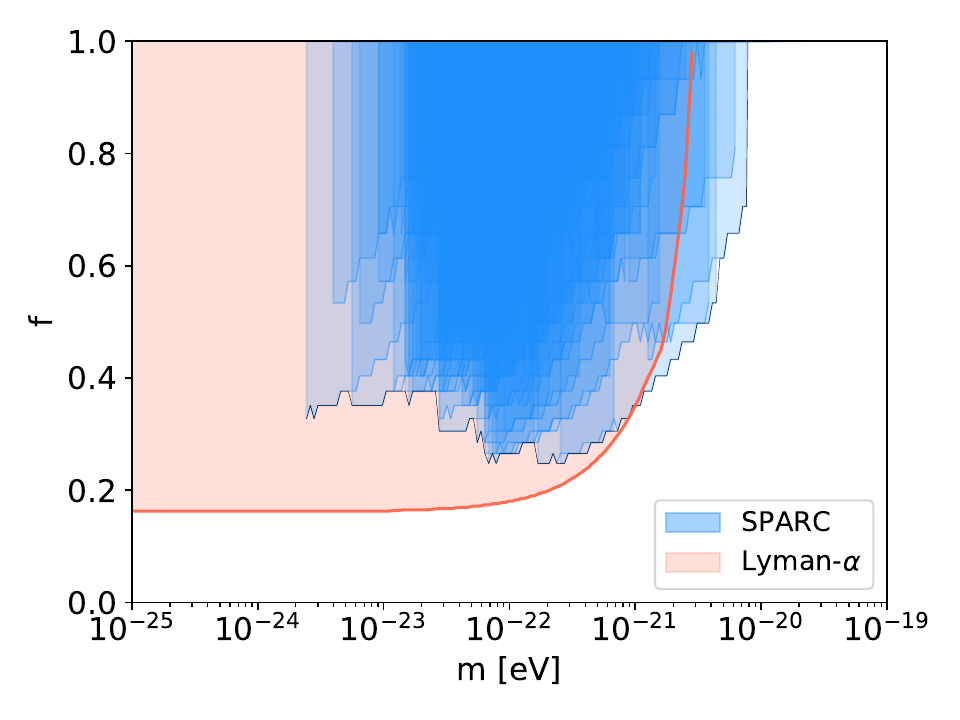}
  \includegraphics[width=.45\textwidth]{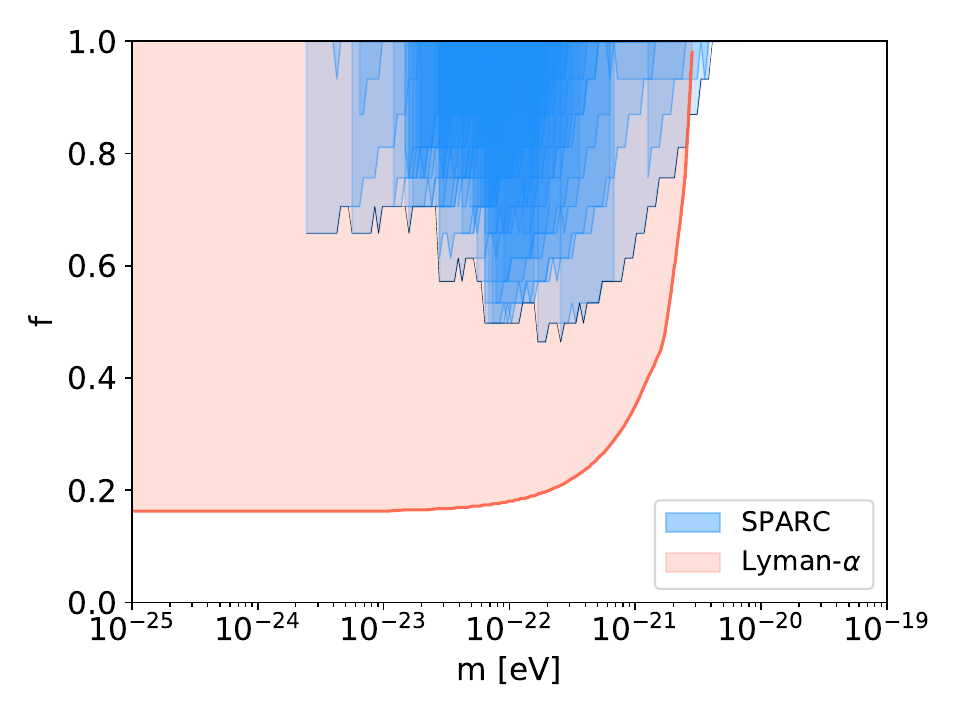}
  \caption{Cosmological ULDM fraction $f$, disfavored by combining SPARC data together with the condition of sufficiently fast dynamical relaxation, for a range of ULDM particle mass $m$. For the relaxation criterion, in the \textbf{left} panel we impose $ r_{\rm relax} > r_{\rm supply} $ ({\it c.f.} Eq.~\eqref{eq:SHcrit}). The \textbf{right} panel tests a stricter version of the criterion, with $ r_{\rm relax} > 2r_{\rm supply} $. We put bounds on $f$ in the mass range where the rotation curve data is in tension with the soliton-halo relation at 95\% confidence level. In the \textbf{top} panel we impose $M_{\rm sol}< M_{\rm SH}$  ($m$ range below the dashed line in Fig.~\ref{fig:final-result-intro}); in the \textbf{bottom} panel we impose $M_{\rm sol}< 0.5\;M_{\rm SH}$ ($m$ range below the whole red band in Fig.~\ref{fig:final-result-intro}). 
        }
	\label{fig:f-m-ULDM}
\end{figure*}

Using the criterion Eq.~\eqref{eq:SHcrit}, we can translate the observational constraints of Sec.~\ref{ssec:data} into constraints on the ULDM fraction $f$. For each value of $m$, we scrutinize the SPARC database and find the smallest value of $f$ for which: (i) solitons with a mass $M_{\rm sol}=M_{\rm SH}$ are in tension with the data to some specified statistical significance, as in \reffig{final-plot}; (ii) the condition Eq.~\eqref{eq:SHcrit} is satisfied, for all of the galaxies that yield this tension. 
We show the result of this exercise in \reffig{f-m-ULDM}, where we also explore the sensitivity of our results to the details of the relaxation condition. The left panels show the disfavored range of $ f $ versus $ m $, using Eq.~\eqref{eq:SHcrit}. The right panels use a stricter condition, $ r_{\rm relax} > 2r_{\rm supply} $. In the upper panels we take the soliton-halo relation as in tension with data when $M_{\rm sol} < M_{\rm SH}$ at 95\% confidence level, while in the lower panels we use a stricter criterion $M_{\rm sol} < 0.5\,M_{\rm SH}$.
For comparison, we also display the cosmological Lyman-$ \alpha $ constraints.
%


\section{Additional comments}\label{s:com}
\subsection{More than one $m$}
\label{ssec:multim}
If one species of ULDM exists, there may just as well be more than one~\cite{Svrcek:2006yi}; indeed, this could be the expectation in scenarios such as that advocated in~\cite{Arvanitaki:2009fg}. It is, therefore, quite relevant to check if the constraints we derived so far could become weakened by the presence of additional species of ULDM. We try to address this question in this section. Our analysis suggests that the constraints derived under the assumption of only one species of ULDM are, in fact, more likely to become even tighter, if additional species exist. Moreover, additional, even subdominant species of ULDM could open up new regions of the parameter space for which observational imprints in galaxy kinematics could be sought after. The reason this happens is dynamical relaxation, which could become more efficient with additional ULDM components.

If more than one species of ULDM exists, then quasiparticles of one species should also induce dynamical relaxation on the other species. We can define the relaxation time $ \tau_{ij} $ of species $i$ due to the gravitational interaction with $j$. Estimating the relaxation process as coming from two-body encounters between ULDM quasiparticles~\cite{Hui:2016ltb,Bar-Or:2018pxz} shows that $ \tau_{ij}\approx \tau_{jj} $ (see Appendix~\ref{app:relax-mult-axions} for a derivation). Note that $ \tau_{ii}$ is given by Eq.~\eqref{eq:relax-time-ULDM}, with $\rho\to f_i\rho$ and $m\to m_i$.
The effective relaxation time of a species should, thus, be given by $ \tau_{i} = (\sum_j \tau_{ij}^{-1})^{-1} $. For example, if two dominant species of ULDM exist in the system, we can estimate the effective relaxation time of species 1 as (taking $ b\sqrt{2}\approx 1 $)
\begin{align}
\label{eq:tau1multi}
\tau_1 \approx \frac{1}{12\pi^3}\frac{m_1^3\sigma^6}{G^2 \rho^2}\frac{1}{f_1^2\ln\Lambda_1}\left[1+\frac{\left({f_2}/{f_1}\right)^2}{\left({m_2}/{m_1}\right)^3X^2}\right]^{-1} ,
\end{align}
where we wrote the Coulomb logarithm $X=\left({\ln\Lambda_1}/{\ln\Lambda_2}\right)$ and $ \ln\Lambda_i=\ln\Lambda_1+\ln(m_i/m_1)$.  

\onecolumngrid
\twocolumngrid

Up to the logarithmic correction, the relaxation time due to species $i$ scales as $m_i^3/f_i^2$. This means that the presence of even a small amount ($f_2\ll1$) of ``spectator" ULDM with a very small $m_2$ could, in principle, dominate the relaxation process for another, potentially dominant ($f_1\sim1$) ULDM species, if $(m_1/m_2)^3>(f_1/f_2)^2$. What cuts off this potential enhancement of relaxation is the Coulomb log: Eq.~\eqref{eq:relax-time-ULDM} should break down for $\ln\Lambda\lesssim1$. Thus the effect can take place only as long as $m_2\gg1/(\sigma R)\approx4\times10^{-23}\left(\frac{{\rm 1~kpc}}{R}\right)\left(\frac{50~{\rm km/s}}{\sigma}\right)$~eV. 

As an aside, note that in a multispecies scenario of axionlike particles, the cosmological relic abundance of each species is expected in the minimal vacuum misalignment mechanism to satisfy $ \Omega_i = f_i\Omega_{\rm DM} \propto \theta_i^2F_i^2 m_i^{1/2} $, where $ F_i $ is the axion decay constant and $\theta_i$ is a vacuum misalignment angle (expected to be of the order of unity for initial conditions set before inflation). Assuming that $ F_i\sim F $ is roughly universal among the different species, and neglecting $\mathcal{O}(1)$ differences in initial  misalignment angles, we find the parametric dependence of the factor in the parentheses in \refeq{tau1multi}
\be
\left[\sum_j \frac{(f_j/f_1)^2}{(m_j/m_1)^3}\right]^{-1} \sim \left[\sum_j \frac{m_1^2}{m_j^2}\right]^{-1} \sim \frac{{\rm min}(m)^2}{m_1^2} \;.
\ee
The species participating in the sum are those for which $m_i\gg1/(\sigma R)$. Even with this condition, it is possible in principle for this factor to enhance the efficiency of dynamical relaxation, compared to naive expectations with a single species of ULDM. 

Suppose there is one species of ULDM with $ m_1 $, $ f_1 $, and a second species with $ m_2 < m_1 $ and $ f_2 $. This setup could lead to stronger constraints on $ f_1 $, compared to the single-species scenario. Figure~\ref{f1-m1-fixed-f2-m2} demonstrates this point. The region inside the gray-colored contour corresponds to single-species relaxation discussed previously. The total blue-shaded region is the constraint on $ f_1 $ versus $ m_1 $ that would be obtained if, in the relaxation time computation, we include an additional species of ULDM at $m_2= 10^{-23} $~eV and $f_2=0.1$, consistent with the Lyman-$ \alpha $ limit. 
%
\begin{figure}[t]
  \includegraphics[width=.43\textwidth]{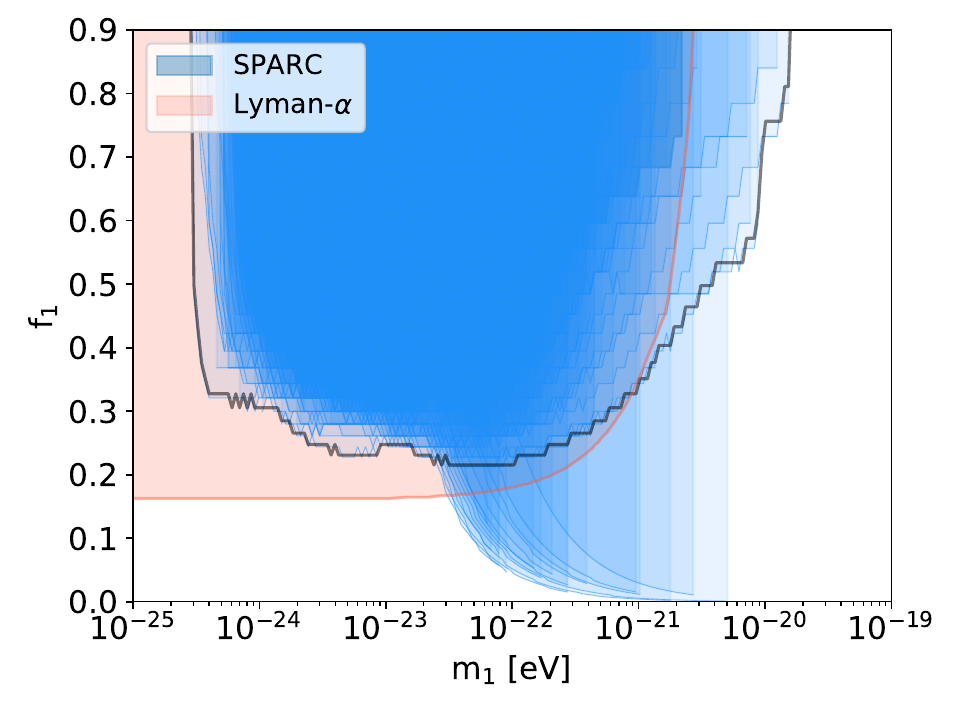}
  \caption{Demonstration of the impact of a light ``spectator" ULDM species with mass $m_2= 10^{-23}$~eV and DM fraction $f_2=0.1$ on the dynamical relaxation of a second species with $m_1$. The region inside the gray contour shows the constraint derived in the $m_1,f_1$ plane, neglecting the impact of the $m_2$ species. The all blue shaded region shows the constraint when the $m_2$ species is accounted for in the relaxation time estimate. The sharp cutoff at the right edge of the blue region for each galaxy is due to the soliton-halo relation becoming compatible with the data at high $m_1$.  
}
	\label{fig:f1-m1-fixed-f2-m2}
\end{figure}

\subsection{Soliton random walk}\label{ss:randwalk}
Throughout our analysis we considered the soliton to be at rest at the bottom of the host halo gravitational potential well. Simulations in Ref.~\cite{Schive:2019rrw} (see also discussion in Refs.~\cite{Li:2020ryg,Chowdhury:2021zik,Chiang:2021uvt}) suggest that, instead, the soliton may be constantly moving in a random walk at the center of the halo. For the halo studied in~\cite{Schive:2019rrw}, which was intended to mimic the dwarf galaxy Eridanus II with a virial mass of the order of $\sim10^{10}$~M$_\odot$, the range of the soliton motion was found to be of the same order as the soliton core radius, with a timescale of the order of the gravitational dynamical timescale. 

Soliton random walk~\cite{Schive:2019rrw} could affect our constraints, because it would induce a time-varying potential. Dedicated simulations would be needed to conclusively check the effect, and we think that our results strongly motivate such dedicated simulations. This said, we suspect that it is unlikely to ameliorate our bounds significantly. The first point to make is that the benchmark soliton-halo relation is in a rather significant tension with respect to many rotation curves. Judging from Figs.~\ref{fig:final-result-intro} and~\ref{fig:final-plot}, even at 10$\sigma$~C.L., with a conservative treatment of the background mass profile of galaxies, the soliton-halo relation overpredicts the rotation velocity of many galaxies by factors of a few.\footnote{A factor of $\sim3$ overprediction of the velocity comes from a factor of $\sim10$ overprediction of the central mass of the halo, which is what the $y$ axis of Fig.~\ref{fig:final-plot} shows.}
The soliton-induced ``bump" in the rotation velocity of a star peaks at a radius $x_{\rm peak}\approx2x_{\rm c}$ (see, e.g. \BarA), where $x_{\rm c}$ is the core radius as defined in Refs.~\cite{Schive:2014dra,Schive:2014hza}. Displacing the soliton by $2x_{\rm c}$ would decrease the soliton-induced rotation velocity at its former peak position by only about 40\%, compared to the factor of a few mismatch noted above. The second point is that soliton random walk in 3D in the central region of a cold stellar disk, like those of some low-surface-brightness, low-dispersion galaxies in SPARC~\cite{Lelli:2016zqa} (see discussion in~\BarB), is likely to heat up and disperse such cold disks, analogously to the effect found in Ref.~\cite{Schive:2019rrw} when considering the nuclear cluster of Eridanus II. Investigating this effect further is beyond our present scope, but we suspect that it may amplify, rather than ameliorate, the tension for ULDM in disk galaxies.

\section{Conclusions}
\label{sec:conclusions}

We used galaxy rotation curves to look for and constrain ultralight dark matter, following and extending earlier work by \BarA and \BarB. The analysis is independent from and complementary to cosmological bounds in the literature. As already shown in \BarA and \BarB, the soliton-halo relation found in simulations is strongly disfavored by the data in the range of ULDM particle mass around $m\sim10^{-22}$~eV, where it was directly tested in numerical experiments. Here we have shown that the data disfavor the soliton-halo relation over a broad range, $10^{-24} \; \mathrm{eV}< m < 2\times 10^{-20}\;\mathrm{eV}$. In much of this range, the relation was not directly tested numerically; however, theoretical analysis of soliton formation via gravitational dynamical relaxation suggests that in many galaxies, a soliton adhering to the soliton-halo relation should indeed form. While turning this argument into a robust constraint would require  dedicated simulations, we believe that it (i) provides adequate motivation for the search in the data, and (ii) having done the search, the lack of significant soliton features disfavors (if indeed not robustly excludes) ULDM in a broad range of $m$.  

As an aside, we argued that the presence of multiple species of ULDM, as might be expected in the string axiverse scenario, could lead to dynamical relaxation becoming more efficient than would be naively estimated in case the ULDM makes up just a fraction $f<1$ of the total cosmological DM. This suggests that having ``more than one $m$" could open up unexpected regions in parameter space where the signature of an ULDM soliton might be meaningfully sought after.

\acknowledgments
N.B. is grateful for the support of the Clore scholarship of the Clore Israel Foundation. K.B. and N.B. were supported by Grant No. 1784/20 from the Israel Science Foundation. C.S. is supported by the Foreign Postdoctoral Fellowship Program of the Israel Academy of Sciences and Humanities, partly by the European Research Council (ERC) under the EU Horizon 2020 Program (ERC-CoG-2015 - Proposal n.~682676 LDMThExp), and partly by Israel Science Foundation (Grant No.~1302/19). We thank the hospitality of INFN Galileo Galilei Institute for Theoretical Physics, where participation in the workshop `New Physics from The Sky' caused our publication to be delayed by approximately two weeks (because there were so many other interesting topics to think about).

\begin{appendix}

\section{modeling a soliton with realistic background profiles}
\label{app:modelsAB}
The constraints we considered in Sec.~\ref{ssec:data} and most of the main text were based on a conservative analysis, in which no attempt was made to fit the actual rotation curve data, and an ULDM soliton was disfavored only if the soliton-induced rotation velocity by itself overshoots the data. In reality, of course, we expect additional contributions to the rotation curve, coming from baryonic matter as well as from ULDM outside of the soliton core, or perhaps non-ULDM components of DM in scenarios in which $f<1$. The goal of this appendix is to estimate the impact of such additional mass components on the analysis.

Regarding the baryonic mass, the SPARC database~\cite{Lelli:2016zqa} includes model estimates of the baryon-induced velocity components, with radial profiles anchored to stellar ($3.6\mu$) surface brightness and HI column density data:
\begin{align}
V_{\rm bar}^2(r_i)
& =
|V_{\rm gas}(r_i)|V_{\rm gas}(r_i) + \Upsilon_{\rm disk}     |V_{\rm disk}(r_i)|V_{\rm disk}(r_i)\no\\ & + \Upsilon_{\rm bulge} |V_{\rm bulge}(r_i)| V_{\rm bulge}(r_i). \label{eq:8}
\end{align}
We allow the mass-to-light ratios $ \Upsilon_{\rm disk,bulge} $ to vary in the fit. The gas component is held fixed as given in SPARC\footnote{\BarB did an independent gas model reconstruction for a few sample galaxies, arriving at similar results to those reported in~\cite{Lelli:2016zqa}.}.

We will consider two models for the DM or ULDM outside of the soliton region: an NFW profile \cite{Navarro:1996gj}, and a Burkert profile \cite{Burkert:1995yz}. 

\subsubsection{Soliton + NFW}
\label{sec:soliton-+-nfw}
In addition to the baryonic contributions and the soliton core, this model includes an NFW density profile:
\begin{align}
\label{eq:NFW-rho1}
\rho_{\rm NFW}(r)
  & =
      \dfrac{\rho_sr_s}{ r \left ( 1 + r/r_s \right )^2}.
\end{align}
$\rho_{\rm NFW}$ has two parameters, which we take to be the NFW radius $r_s$, and the concentration parameter $c$, related to the density parameter $\rho_s$ via $\rho_s = \rho_c (200 c^3)/(3(\ln(1+c) - c/(1+c)))$, with $\rho_c$ the critical density. 

We consider two versions of the model. In the first, we simply add the NFW component in addition to the soliton profile. This way, even in the region where the soliton profile dominates the density, the two DM components overlap. This scenario may be quite relevant, if ULDM makes up just a fraction of the total DM. 

In the second version of the model, we match the density of the soliton and NFW profiles at a transition radius $r_t$, where $\rho_{\rm sol}(r_t) = \rho_{\rm NFW}(r_t)$, and consider the NFW (soliton) component only outside (inside) of $r_t$ (in case $\rho_{\rm sol}$ is subdominant everywhere, $r_{t} = 0$, we use only the NFW profile):
\begin{align}
\label{eq:NFW-rho2}
\rho_{\rm DM}(r)
  & =
    \begin{cases}
      \rho_{\rm NFW}(r), & r > r_t\\
      \rho_{\rm sol}(r), & r < r_t
    \end{cases}
\end{align}

In both versions, the total DM mass profile has four free parameters, $M_{\rm sol}, m, c$, and $r_s$. 
We define the total model-predicted velocity as 
\begin{align}
\label{eq:Vth}
V_{\rm th}^2(r; \theta_{\rm th}; \theta_{\nu})
& =
\frac{G M_{\rm DM}(r; \theta_{\rm th}; \theta_\nu)}{r} + V_{\rm bar}^2(r; \theta_{\nu}),
\end{align}
where $\theta_{\rm th}=\{M_{\rm sol}, m\}$ and $\theta_{\nu} = \{c, r_s, \Upsilon_{\rm disk}, \Upsilon_{\rm bulge}\}$. 
This is compared with the observed velocity data using 
\begin{align}
\label{eq:chi2-modAB}
\chi^2(\theta_{\rm th}; \theta_{\nu})
& =
\sum_{i}
\left (
\frac{V_{\rm th}(r_i; \theta_{\rm th}; \theta_{\nu}) - V_{\rm obs}(r_i)}{\Sigma_i}
\right )^2,
\end{align}
where $V_{\rm obs}(r_i)$ and $\Sigma_i$ are the measured rotation velocity and standard deviation in the $i$th radius bin, respectively.  The summation is over the radial data bins. 

We scan a grid of values of the ULDM particle mass $m\in(10^{-24}, 10^{-19})$~eV. For each value of $m$, we allow the remaining model parameters to vary in the following range:
\begin{align*}
\begin{array}{rlrl}
M_{\rm sol}/M_\odot
& \in (10^{4.5}, 10^{12})
& 
& 
\\
\Upsilon_{\rm disk}
& \in (0, 5)
& r_s/\mathrm{kpc}
& \in (5, 30)
\\
\Upsilon_{\rm bulge}
& \in (0, 5)
& c
&      \in (5,30)
\end{array}
\end{align*}
To constrain $M_{\rm sol}$, we perform a log-likelihood ratio test for $M_{\rm sol} \cup \theta_{\nu}$ separately for each value of $m$, minimizing  the $\chi^2$ with respect to $\theta_\nu$.

\subsubsection{Soliton + Burkert}
\label{sec:soliton-+-burkert}
This model is identical to that in the previous section (including the two versions of adding the soliton term), apart from replacing the NFW density profile with the Burkert profile, 
\begin{align}
\label{eq:burkert}
\rho_{\rm Bkt}(r)
& =
\frac{\rho_0}{ \left (1 + \frac{r}{r_0}\right ) \left ( 1 + \left ( \frac{r}{r_0} \right )^2 \right )}. 
\end{align}
We express $\rho_0 = \rho_c \delta_0$, with $\rho_c$ the critical density of the Universe. The total DM mass profile has four free parameters, $M_{\rm sol}, m, \delta_0$, and $r_0$. 
For each value of $m$ on a fixed grid $m\in(10^{-24}, 10^{-19})$~eV, we allow the remaining parameters to vary in the following range:
\begin{align*}
\begin{array}{rlrl}
M_{\rm sol}/M_\odot
& \in (10^{4.5}, 10^{12})
& 
& 
\\
\Upsilon_{\rm disk}
& \in (0, 5)
& \log_{10}(\delta_0)
& \in (-1, 6)
\\
\Upsilon_{\rm bulge}
& \in (0, 5)
& r_0/\mathrm{kpc}
& \in (1, 60)      
\end{array}
\end{align*}

\section{Relaxation of multiple axions}
\label{app:relax-mult-axions}
Reference~\cite{Hui:2016ltb} pointed out that gravitational dynamical relaxation (see, {\it e.g.},~\cite{binney2011galactic} for a textbook review) in an ULDM field can be understood effectively as being mediated by two-body scattering events of massive quasiparticles (QPs). The QPs arise from interference patterns in the field, with a characteristic coherence length of $\lambda_{\rm dB}\sim2\pi/(m\sigma)$. If the ULDM ambient density is $\rho$, the mass of each QP is of the order of $M_{\rm QP}\sim(4\pi/3)\rho\lambda_{\rm dB}^3\sim 6\times10^9\left(\rho/0.1~{M_{\odot}{\rm pc}^{-3}}\right)\left(50~{\rm kms^{-1}}/\sigma\right)^3\left(10^{-22}~{\rm eV}/m\right)^3~M_{\odot}$. The effective QP description was made rigorous in analytical studies~\cite{Bar-Or:2018pxz} and further elucidated and calibrated in numerical simulations~\cite{Levkov:2018kau}.

Consider the case of just one species of ULDM, with particle mass $m$, ambient density $\rho$, and QP mass $M_{\rm QP}$, and consider the motion of a single test particle (not necessarily ULDM), with mass $m_{\rm test}\ll M_{\rm QP}$, traversing this medium. The mean time between significant collisions of the test particle against QPs in the background is (ignoring order unity factors)
\be\label{eq:tauest}\tau\sim\frac{1}{n_{\rm QP}\sigma b^2},\ee
where $n_{\rm QP}=\rho/M_{\rm QP}$ is the QP number density, $\sigma$ is the velocity dispersion in the system (pertaining to the QPs and to the test particle alike), and $b$ is the impact parameter for a significant collision. We define significant collisions as collisions that change the velocity of the test particle by an order unity factor; thus,
\be b\sim\frac{GM_{\rm QP}}{\sigma^2}.\ee
Inserting this into \refeq{tauest}, and using the definition of $M_{\rm QP}$, we have
\be\label{eq:tauest1}\tau\sim\frac{m^3\sigma^6}{G^2\rho^2}.\ee
Up to the Coulomb log, \refeq{tauest1} has the same parametric scaling as \refeq{relax-time-ULDM}. Of course, the equations describe the same process; we could just as well have set the test particle mass to $m_{\rm test}=m$, making it part of the ULDM. The numerical factors required to make \refeq{tauest1} precise were calibrated in Refs.~\cite{Bar-Or:2018pxz,Levkov:2018kau}.

Using this understanding, the dynamical relaxation induced by one ``spectator" species of ULDM, with particle mass $m_2$ and DM fraction $f_2$, onto another ULDM species with particle mass $m_1$, is simply obtained from \refeq{relax-time-ULDM}, substituting $\rho\to f_2\rho$ and $m\to m_2$. In the main text we referred to this ``off-diagonal" relaxation time as $\tau_{12}$.

\end{appendix}

\bibliography{bib}
\bibliographystyle{utphys}

\end{document}